# Attosecond Electron Pulses from Interference of

# Above-Threshold de Broglie waves

Authors: **Sándor Varró and Győző Farkas**

Affiliation: Research Institute for Solid State Physics and Optics

of the Hungarian Academy of Sciences

Letters: H-1525 Budapest, PO Box 49, Hungary

Phone: +36-1-392-2635, Fax: +36-1-392-2215

E-mails: varro@sunsev.kfki.hu  and  farkas@szfki.hu

Short title: Attosecond Electron Pulses

Number of pages: 40

Number of figures: 6

Title: **Attosecond Electron Pulses from Interference of Above-Threshold de Broglie waves**

Authors: **Sándor Varró and Győző Farkas**

**Abstract.** It is shown that the above-threshold electron de Broglie waves, generated by an intense laser pulse at a metal surface, are interfering to yield attosecond electron pulses. This interference of the de Broglie waves is an analogon of the superposition of high harmonics generated from rare gas atoms, resulting in trains of attosecond light pulses. Our model is based on the Floquet analysis of the inelastic electron scattering on the oscillating double-layer potential generated by the incoming laser field of long duration at the metal surface. Owing to the inherent kinematic dispersion, the propagation of attosecond de Broglie waves in vacuum is very different from that of attosecond light pulses, which propagate without changing shape. The clean attosecond structure of the current at the immediate vicinity of the metal surface is largely degraded due to the propagation, but it partially recovers at certain distances from the surface. Accordingly, above the metal surface, there exist "collaps bands", where the electron current is erratic or noise-like, and there exist "revival layers", where the electron current consist of ultrashort pulses of about 250 attosecond durations in the parameter range we considered. The maximum value of the current densities of such ultrashort electron pulses has been estimated to be of order of couple of tenth of $mA/cm^2$. The attosecond structure of the electron photocurrent can, perhaps be used for monitoring of ultrafast relaxation processes in single atoms or in condensed matter.

## 1. Introduction

The generation of attosecond (1 $as$ = $10^{-18}$ s) light pulses have already long been proposed by Farkas & Tóth (1992). The first experimental indication of attosecond localization in time of the *high-harmonic signal* stemming from the non-linear response of rare gas jets excited by high-intensity laser pulses was published by Papadogiannis et al. (1999). The generation of attosecond light pulses is in complete analogy of mode-locking producing picosecond pulse trains through the Fourier synthesis of the laser cavity eigenmodes. Because the spacing of the high-harmonic components (namely the optical frequency $\omega=2\pi\nu$) is much larger than the spacing of a the cavity eigenmodes (of order of one GHz) of a usual laser, the Fourier synthesis results in a series of spikes whose width can be much smaller than the optical period (which is roughly of order of $10^{-15}$ s), i.e. we receive here a train of sub-femtosecond (attosecond) pulses. At the beginning it was questionable whether the phases of the high-harmonic components are really locked (i.e. the difference of the phases is a smooth (possibly a constant) function of the harmonic order, which is crucial to have constructive interference), but in the meantime many thorough analysis have been carried out, e.g. by L'Huillier et al. (1992), Lewenstein et al. (1994), Antoine et al. (1996) and Saliéres (2001), and it has become clear that the phase-locking usually automatically takes place due to the generating mechanism itself. Paul et al. (2001) have measured first an *attosecond pulse train* with sub-pulses of 250 $as$ duration, and Tzallas et al. (2003) repeated a refined version of the earlier experiments by Papadogiannis et al. (1999) for 780 $as$ pulse trains. López-Martens et al. (2005) have produced by now the cleanest attosecond pulse trains of 170 $as$ generated rutinely. Christov et al. (1997) have predicted theoretically the appearance of a *single attosecond pulse*, and such a pulse of duration 650 $as$ were first produced by Hentschel et al. (2001). Later Sansone et al. (2006) have produced one single 120 $as$ pulse by using an optical gate, and it has been predicted by Tsakiris et al. (2006), that from the cut-off region of the high-harmonic spectrum of an overdense plasma layer, where the spectrum is essentially a quasi-continuum, isolated single attosecond light pulses can be obtained.

We note here that recently there has been much labour put into the classical simulations of various processes (generation of coherent x-rays, laser acceleration of electrons) in laser-plasma interaction (see e.g. Pukhov & Meyer-ter-Vehn, 2003, Kiselev et al., 2004, Quèrè et al., 2006 and Tsakiris et al., 2006). Moreover, the first experimental results by Hidding et al. (2006) appeared on the generation of quasi-monoenergetic electron bunches by strong laser fields. The wide range of applicability of ultrashort laser pulses have been recently also represented in the present journal by several contributions. For instance, Eliezer et al. (2005) have reported on the production by femtosecond laser pulses of crystal nanoparticles for aluminium and nanotubes for carbon on a transparent heat-insulating glass substrate. Kanapathipillai (2006) have worked out a nonlinear oscillator model to describe the nonlinear absorption of ultrashort laser pulses by clusters. Sherlock et al. (2006) have shown by a numerical study that it is necessary to take into account the collisional heat transport into the target in order to correctly model the absoption rate of laser pulses of duration of 100 fs, and of intensities of order $10^{15}$ W/cm$^2$ at the front of the target surface. According to the theoretical studies on the interaction of a short laser pulse with metals performed by Anwar et al (2006), the laser-induced electric field inside the target is responsible for an induction of the current density, which causes, after all, electronic heat conduction. Laser-induced acceleration and manipulation of high-energy charged particle beams are still subjects of extensive theoretical and experimental research. For instance, Lifschitz et al. (2006) have recently proposed a new scheme for a compact GeV laser plasma accelerator. According to the simulations performed by these authors their method would yield the production of high quality, monoenergetic and sub-50 fs electron bunches at the GeV energy level. Willi et al.

(2007) have proposed a novel technique for focusing and energy selection of high-current MeV proton beams. In their scheme the transient electrostatic field induced by an ultra-short laser pulse is resposible for the "micro-lensing", i.e. for the focusing and for the selection of a narrow band out of the broadband spectrum of protons generated from a separate laser-irradiated thin foil target. The distortion of the Fermi distribution and the step-like occupation of the energy levels of the electrons due to the influence of a 100 fs strong laser pulse have been recently studied by Schwengelbeck et al. (2002) on the basis of a time-dependent exact quantum-mechanical analysis of a many-electron system. Besides these investigations of the interaction of ultra-short but "non-attosecond" laser pulses with matter, there has been a wide experimental and theoretical research carried out concerning attosecond electron pulses, too. Lindner et al. (2005) performed attosecond electron double-slit experiments, and Johnsson et al. (2005) studied electron wave packet dynamics in strong laser fields. A time-dependent calculation has been carried out by Mauritsson et al. (2005) in order to study the properties of electron wave packets generated by attosecond laser pulse trains. In this context see also the works by Remetter et al. (2006) on attosecond wave packet interferomety, Breidbach et al. (2005) on the attosecond response to the removal of an electron, Niikura et al. (2005) on attosecond electron wave packet motion, Hu et al. (2006) on the possibility of attosecond pump-probe experiments for exploring the ultrafast electron motion inside an atom, and Fill et al. (2006) on sub-fs electron pulses for ultrafast electron diffraction. Quite recently attosecond real-time observation of electron tunneling in atoms has been reported by Uiberacker et al. (2007).

The *above-threshold electron spectra* of nonlinear photoionization induced by *relatively long laser pulses*, analysed thoroughly e.g. by Agostini (2001), Paulus and Walther (2001) and recently by Banfi et al (2005), have common features with the corresponding high-harmonic spectra. The initial fall-off, the (occasionally rising) plateau and the sharp cut-off are present in each cases. In case of multiphoton photoelectric effect of metals Farkas & Tóth (1990) and Farkas et al. (1998) measured very high-order above-threshold electrons coming from metal targets. The theoretical interpretation of these results has been given by one of the present authors in Varró & Ehlotzky (1998) and recently in Kroó et al. (2007), on the basis of the so-called laser-induced oscillating double-layer potential model. This model is based on a Floquet-type analysis of the inelastic electron scattering on the oscillating double-layer potential generated by the incoming laser field at the metal surface. The model has already been succesfully used to interpret the experimental results on very high order surface photoelectric effect in the near infrared (Farkas & Tóth, 1990) and in the far infrared regime (Farkas et al., 1998). By analogy, one may think that if the phases of the *above-threshold electron de Broglie waves* generated at the metal surface are locked (i.e. the difference of the phases of the neighbouring components is a smooth, possibly a constant function of the order, namely the number of absorbed photons), then the Fourier synthesis of these components yields an *attosecond electron pulse train* emanating perpendicularly from the metal surface, quite similarly to the generation of attosecond light pulses from high harmonics (which, on the other hand, are propagating in the specular direction). This expectation is quite natural, because the spacing of the electron peaks in the frequency space is just the optical frequency $h\nu/h=\nu$, like in the case of high-harmonic generation.

We emphasize that, owing to the inherent kinematic dispersion, the propagation of the attosecond de Broglie waves in vacuum is very different from that of attosecond light pulses, since the rest mass of the electrons is not zero, in contrast to that of the photons. The clean attosecond structure at the immediate vicinity of the metal surface is largely spoiled due to the propagation, even in vacuum, but it partially recovers at certain distances from the surface. On the basis of the existence of a plateau in the electron spectrum too, as has been discussed e.g. by Agostini (2001), Paulus and Walther (2001) and Kystra et al. (2001), we conjecture that a similar effect of interference of de Broglie waves may exist in the case of the widely studied above-threshold ionization of atoms.

In the present paper in Section 2 we briefly describe the construction of the laser-induced oscillating double-layer potential (energy) of a test electron scattered by the metal surface, and derive the basic wave function matching equation of the the scattered electron. Moreover, we give an approximate analytic expression for the multiphoton scattering amplitudes valid for large final electron energies. This model serves as our basis for studying the high-order multiphoton photoelectric effect. In Section 3 we present the results of the numerical solutions of the matching equations for the scattering amplitudes corresponding to the elementary $n$-photon absorption of the test electron. The time-averaged above-threshold current components and the phases of the multiphoton transmission amplitudes will be discussed. In Section 4 we study the detailed temporal behaviour of the total transmitted current which results from the superposition of the above-threshold de Broglie waves. It will turn out that the ideal *attosecond electron pulse train* appearing at the immediate vicinity of the metal surface collapses to an almost noise-like signal by the inherent kinematic dispersion of the electron waves as they perpendicularly propagate from the surface. On the other hand, there are regions quite far from the surface (even at macroscopic distances) where the clean attosecond structure of the electron current revives. Accordingly, in propagation of the electron signal from the metal surface there appear consecutively "*collapse bands*" and "*revival layers*" of a few nanometer thickness. In Section 5 a short summary closes our paper.

## 2. Laser-induced oscillating double-layer potential at the metal surface and the electron's wave function matching

In the present section, in order to illustrate the appearance of enhanced nonlinearities in the above-threshold electron excitations due to the enlarged electric field of surface inside the metal, let us first calculate the electron displacements in the bulk of the metal caused by the $z$-component of the penetrating electric field $F(x', z', t) = F_0 \exp(z'/\delta) \sin(\omega t - kx')$. Here $F_0$ and $\omega = 2\pi\nu$ are the peak field strength and the circular frequency of the laser, $\delta = 1/k_{metal} \approx c/\omega_p$ is the skin depth and $k = (\omega/c)\sin\theta$ denotes the plasmon wave number, respectively. Moreover, we have introduced the plasma frequency $\omega_p = \sqrt{(4\pi n_e e^2/m)}$, where $e$ and $m$ are the electron's charge and mass, respectively, and $n_e$ denotes the free electron density in the metal. The displacement $\xi(x', z', t) = \alpha_0 \exp(z'/\delta) \sin(\omega t - kx')$ of an electron in the bulk, at an average position $(x', y', z')$ can be obtained from the solution of the corresponding Newton equation, where the amplitude of oscillation is given as $\alpha_0 = eF_0/m\omega^2$. Here we have taken into account parametrically the $z'$ and $x'$ dependence of the penetrating electric field, which is a justified approximation at relatively moderate incoming laser intensities to be cosidered below. The potential energy $U_d(\boldsymbol{x}, t; \boldsymbol{x'})$ of a test electron at position $\boldsymbol{x} = (x, y, z)$ in the joint Coulomb field of a background ionic core at a fixed position $\boldsymbol{x'} = (x', y', z')$ and of an associated oscillating background electron, is given by $U_d(\boldsymbol{x}, t; \boldsymbol{x'}) = e^2/|\bar{x} - \bar{x}'(t)| - e^2/|\bar{x} - \bar{x}'|$. Here $\bar{x}'(t) = \bar{x}' + \bar{\varepsilon}_z \xi(x', z', t)$ is the instantaneous position of the oscillating background electron, with $\varepsilon_z$ being a unit vector perpendicular to the metal-vacuum interface, pointing to the positive $z$-direction. The total potential energy of a test electron is the sum of all the contributions coming from the interactions originating at the positions $\boldsymbol{x'}$, i.e.

$$U_d(\bar{x}, t) = \sum_{\bar{x}'} U_d(\bar{x}, t; \bar{x}') \to n_e e^2 \int d^3 x' \frac{\xi(x', z', t)(z - z')}{|\bar{x} - \bar{x}'|^3} + O(\xi^2) \, . \tag{1}$$

In obtaining Eq. (1) we have used the continuum limit of the summation and we have expanded the joint Coulomb interaction of the background in powers of the oscillating displacement $\xi$. The first term on the right hand side of Eq. (1) can be calculated analytically,

$$U_d = 2\pi n_e e^2 \sin(\omega t - kx) \int_{-\infty}^{0} dz' e^{z'/\delta} e^{-k|z-z'|} \frac{z-z'}{|z-z'|} =$$

$$= U_D \sin(\omega t - kx) \begin{cases} \dfrac{2e^{z/\delta}}{1-k^2\delta^2} - \dfrac{e^{kz}}{1-k\delta} & , \ (z<0) \\ \dfrac{e^{-kz}}{1+k\delta} & , \ (z>0) \end{cases} , \qquad (2)$$

$$U_D \equiv (\omega_p / 2\omega)^2 (\delta/\lambda) \mu (2mc^2) , \quad \mu \equiv eF_0 / mc\omega = 10^{-9} I^{1/2} / E_{ph} , \quad \mu^2 = 10^{-18} I\lambda^2 . \qquad (3)$$

In Eq. (3) we have introduced the amplitude $U_D$ of the oscillating collective potential energy of the test electron, Eq. (2), which can take on very large values even for relatively moderate laser intensities (notice the factor $2mc^2$ being just the pair-creation energy $\approx 10^6$ eV). The dimensionless intensity parameter $\mu$ usually shows up in any strong field calculation, its magnitude governes the nonlinearity of the direct laser-electron interactions. In Eq. (3) $I$ denotes the peak laser intensity in W/cm², $E_{ph}$ is the photon energy measured in eV and $\lambda$ is the central wavelength in microns ($10^{-4}$cm). We note that the gradient of the potential energy, Eq. (2), essentially equals to the force acting on a test electron due to a corresponding the collective electric field. The potential given by Eq. (2) looks very much similar to the so-called surface plasmon polariton potential discussed recently in details e.g. by Zayats et al. (2005) in the context of nano-optics. In dipole approximation, i.e. for $|kz|$ and $|kx| \ll 1$, the potential energy in Eq. (2) for relatively moderate laser intensity can be well approximated by the following *double-layer potential energy* $U_d \approx sign(z) U_D \sin(\omega t)$, where $sign(z>0)=1$ and $sign(z<0)=-1$, thus the maximum total jump in the energy equals $2U_D$. Because of this property, henceforth we will call $U_d$ "*laser-induced oscillating double-layer potential*", and describe the scattering of a Sommerfeld electron on this ideal non-conservative potential with the technique of wave function matching. At this point we have to mention that normal electric field strenghts $F_0$ in the vacuum undergoes a reduction of order $(\omega/\omega_p)^2$ by crossing the metal-vacuum interface, but, on the other hand, if we repeat the above same procedure by using the force term $-\partial U_d / \partial z$ in the Newton equation of an oscillating backround electron, in this second iteration we receive essentially the same expression summarized in Eqs. (1), (2) and (3) with the original field strenght $F_0$. In this way the reduction factor is completely compensated by the enhancement factor $(\omega_p / \omega)^2$ due to the collective surface plasmon polaritons. So in this "second iteration" the value of $U_D$ is essentially the same as is given by Eq. (3) with the original vacuum amplitude of field strength $F_0$. The next iteration would be meaningless, because then the amplitude of oscillation of the background electrons $\alpha_0$ would be comparable with the skin depth $\delta$, hence the parametric substitution in Newton's equation could be justified. These detailes will not be discussed any further in the present paper, so henceforth we are planning to use the idealized laser-induced oscillating double-layer potential (energy) in dipole approximation $U_d \approx sign(z) U_D \sin(\omega t)$ throughout the paper, where $U_D$ is defined in Eq. (3).

The concept of the laser-induced oscillating double-layer potential (energy) outlined above has been first introduced in our earlier study (Varró & Ehlotzky, 1998) in order to explain a surprising outcome of one of our experiments (Farkas & Tóth, 1990), namely, the appearance of very large (~600 eV) energy photoelectrons induced by Nd:Glass laser radiation ($h\nu \sim 1.17$ eV) at moderate intensities of some 10 GW/cm$^2$. The main problem there was that the very large nominal order of the photon absorption processes corresponding to the experimental results ($n \sim$ 5-600) could not have been deduced even from the usual non-perturbative approach based on Gordon-Volkov states (Kylstra et al., 2001), since the intensity parameter $\mu = eF_0/mc\omega$ was very small, of order of $10^{-4}$ in that case. That time we have realized that instead of $\mu$, another basic dimensionless parameter "$a$" appears in the analysis in a natural way, when we introduce the interaction with the double-layer potential, which builds up due to the coherent collective excitation of all the electrons within the skin depth. The parameter *a* is defined as $a = 2U_D/h\nu$, where $U_D \approx (\omega_p/4\omega)\mu(2mc^2)$ is the amplitude of the double layer potential energy of a test electron. The size of this *a* governing the degree of nonlinearity turned out to be just of order of 500 for the mentioned experiment, hence we were able to explain the basic features of the measured electron spectra. In the meantime we have applied the same method (Kroó et al., 2007) for the theoretical interpretation of another strange experimental results (Farkas et al., 1998) concerning electron emission from gold cathodes (work function ~4.7 eV) irradiated by mid-infrared radiation (generated by the Orsay Free Electron Laser) of wavelength up to 12 μm ( $h\nu \sim 0.1$ eV ) in the I ~ 10 MW/cm$^2$ intensity regime. The intensity parameter is extremely small in this case: $\mu \approx 3\times10^{-5}$. The minimum number of photons required for the deliberation of an electron from the binding is of order of 50. As was pointed out by the authors of this paper, both the tunneling model and the multiphoton model predict results many (about 200) orders off the experimental figures.

Now let us turn to the Schrödinger equation of an electron under the joint action of the Sommerfeld step-potential of depth $V_0$ and the double-layer potential derived above. We restrict our analysis to a one-dimensional scattering in dipole approximation, as in our earlier study (Varró & Ehlotzky, 1998), which is justified in case of incoming laser intensities of orders < $10^9$-$10^{11}$ W/cm$^2$ for Ti:Sa lasers, which we have been using in our numerical calculations. Further, to simplify the following analysis we shall take the asymptotic amplitude of the double-layer potential for $z \to -\infty$ as $-U_D$ and for $z \to +\infty$ as $+U_D$, thus we get an idealized double-layer potential that oscillates at circular frequency $\omega$ between $-U_D$ and $+U_D$ at a phase difference $\pi$ between $z > 0$ and $z < 0$. For the sake of completeness of the present paper, in the following few lines we outline the basic equations presented already in Varró & Ehlotzky (1998). The wave function of an electron will then obey the two Schrödinger equations

$$(\hat{p}^2 / 2m - V_0 - U_D \sin \omega t)\Psi_I = i\hbar \partial_t \Psi_I \qquad (z < 0), \qquad (4a)$$

$$(\hat{p}^2 / 2m + U_D \sin \omega t)\Psi_{II} = i\hbar \partial_t \Psi_{II} \qquad (z > 0), \qquad (4b)$$

where the subscript *I* refers to the interior region (metal) and *II* to the exterior region (vacuum), respectively. In order to fulfill the continuity conditions at the metal surface at $z=0$, we make Floquet-type ansätze in terms of the fundamental solutions of Eqs. (4a) and (4b), satisfying the proper mass-shell relations, i.e. the proper free-particle dispersion relations connecting the energies and the corresponding momenta of the reflected and transmitted components,

$$\Psi_I = \left[ \chi_0^{(+)} - \chi_0^{(-)} + \sum_{n=-\infty}^{+\infty} R_n \chi_n^{(-)} \right] \exp\left[ -i \frac{U_D}{\hbar\omega} \cos \omega t \right] \qquad (z < 0), \qquad (5a)$$

$$\Psi_{II} = \sum_{k=-\infty}^{+\infty} T_k \varphi_k^{(+)} \exp\left[ +i \frac{U_D}{\hbar\omega} \cos \omega t \right] \qquad (z > 0), \qquad (5b)$$

where $\chi_n^{(\pm)} = \exp[\pm i q_n z / \hbar - i(E_0 + n\hbar\omega) \cdot t / \hbar]$ with $q_n = [2m(V_0 + E_0 + n\hbar\omega)]^{1/2}$ and, correspondingly, $\varphi_k^{(+)} = \exp[i p_k z / \hbar - i(E_0 + k\hbar\omega) \cdot t / \hbar]$ with $p_k = [2m(E_0 + k\hbar\omega)]^{1/2}$. Notice the opposite signs in the exponential factors in Eqs. (5a) and (5b), which transforms out the interaction with the double-layer potential in Eqs. (4a) and (4b). Here the energy parameter $E_0$ denotes the initial energy of the electron impinging from the inner part of the metal onto the metal-vacuum interface. In the numerical calculations it will be taken as the negative of the work function, i.e. $-A$, which means that we assume that the test electron starts from the Fermi level. For later convenience, we have separated in Eq. (5a) the trivial elastic back-scattered part $\chi_0^{(-)}$ from the total back-scattered wave function. The unknown multiphoton reflection and transmission coefficients $R_n$ and $T_k$, respectively, can be determined from the matching equations, i.e. from the continuity of the wave function, $\Psi_I(0,t) = \Psi_{II}(0,t)$ and of its spatial derivative, $\partial_z \Psi_I(0,t) = \partial_z \Psi_{II}(0,t)$, which relation must hold for arbitrary instants of time. By using the generating functions of the ordinary Bessel functions of first kind $J_n(z)$ of order $n$ (Gradshteyn and Ryzhik, 2000) to Fourier decompose the time-dependent exponentials in Eqs. (5a) and (5b), the matching equations yield the following infinite set of algebraic equations,

$$R_n = \sum_{k=-\infty}^{+\infty} J_{n-k}(a) i^{n-k} T_k , \qquad (6a)$$

$$\delta_{n,0} = \sum_{k=-\infty}^{+\infty} J_{n-k}(a) i^{n-k} \big[ (q_n + p_k)/2q_0 \big] T_k , \qquad (6b)$$

where we have introduced the dimensionless parameter $a$, with the definition

$$a \equiv 2U_D / \hbar\omega , \quad U_D \equiv (\omega_p / 2\omega)^2 (\delta / \lambda) \mu (2mc^2) \approx (\omega_p / 4\omega) \mu (2mc^2) , \qquad (7)$$

and $U_D$ has already been defined in Eq. (3). Thus the dimensionless parameter $a$ given by Eq. (7) is the ratio of the total maximum jump of the oscillating double layer potential (energy) to the photon energy. The last approximate equation is valid if the plasma frequency of the metallic electrons is much larger than the laser frequency, since in this case the skin depth can be approximated as $\delta = c / (\omega_p^2 - \omega^2)^{1/2} \approx c / \omega_p$. The magnitude of the parameter $a$ governes the extension of the kernel matrix in the sets of algebraic equations (6a) and (6b), hence it determines the degree of nonlinearity of the multiphoton excitation.

The time-averaged outgoing electron current components (for which $p_n$ is real), corresponding to $n$-photon absorption, can be obtained from $\Psi_{II}$. We normalize these current components with respect to the incoming current, and get the dimensionless quantities

$$j_t(n) = (p_n / q_0) \cdot |T_n|^2 \qquad (n \geq n_0) , \qquad (8a)$$

where $n_0$ is the minimum number of photons to be absorbed in order to yield true free running outgoing waves, i.e. ionization. The corresponding normalized reflected currents are

$$j_r(n) = (q_n / q_0) \cdot |R_n - \delta_{n,0}|^2 \qquad (n \geq n_1) , \qquad (8b)$$

with a similar meaning for $n_1$ as for $n_0$. The conservation of probability requires $\sum_n \left[ j_t(n) + j_r(n) \right] = 1$, which condition has been used to check the accuracy of the numerical solutions of the matching equations (6a) and (6b).

At the end of the present section we would like to note that for large values of the parameter $a$ given by Eq. (7) an approximate analytic expression has been derived in Varró & Ehlotzky (1998) for the transmission coefficients $T_n$, which are particularly accurate for large values of the multiphoton order $n$. According to this approximation

$$T_n \approx J_n(a)(-i)^n, \quad j_t(n) \approx \left( p_n / q_0 \right) \cdot J_n^2(a) \qquad (a \gg 1) \qquad (n \gg 1) \ . \tag{9}$$

## 3. The time average of the above-threshold current components and the phases of the transmission coefficients. Numerical results

In the present paper we shall use in the numerical calculations the following input parameters for the incoming laser field and for the gold target. We assume a Ti:Sa laser beam which excites the metal surface at grazing incidence of central wavelength $\lambda$ = 800nm, frequency $\omega = 2.36 \times 10^{15}$ Hz, photon energy $h\nu$ = 1.55 eV and of intensity $10^9$ W/cm$^2$, with its polarization being essentially perpendicular to the metal surface. *Moreover, we assume that the pulse duration is much larger than the optical period T = 2.6 femtoseconds, so the carrier-envelope phase effects can be neglected in the present case.* Concerning the carrier-envelope phase effects ("absolute phase effects") in case of interactions with ultrashort laser pulses see e.g. the recent paper by Varró (2007) appearing in the present journal, and the references therein. The depth of the Sommerfeld step potential and the work function of gold are taken $V_0$ = 10.19 eV and $A$ = 4.68 eV, respectively, hence the Fermi energy equals $E_F$ = 5.51 eV. According to Radzig and Smirnov (1985) the electron density of gold is $n_e = 5.9 \times 10^{22}$/cm$^3$, thus the plasma frequency in the bulk equals $\omega_p = 1.37 \times 10^{16}$ Hz >> $\omega$ = $2.36 \times 10^{15}$ Hz. By the definition in Eq. (7), the nonlinearity parameter $a$ equals 60 in the present case. The numerical solution of the infinite set of algebraic equations (6a) and (6b) is accurate up to a fraction of a per cent if we use the truncated kernel in the range of indeces $\{-120 \leq n \leq +120,\ -120 \leq k \leq +120\}$. This can be checked by the conservation of probability displayed after Eq. (8b). We note that the above-mentioned averaging of the multiphoton current means the time-averaging operation $\langle f \rangle = \lim \int_{-T_0}^{+T_0} dt f(t)/2T_0$, as $T_0 \to \infty$, thus, of course, the distribution of $j_t(n)$ and $j_r(n)$ of Eqs. (8a) and (8b) do not reflect back the detailed time behaviour of the total current $(\hbar/m)\,\mathrm{Re}[\Psi^*(-i\nabla)\Psi]$, which will be discussed in the next Section.

At the end of the present Section the time-averaged above-threshold spectrum and the relative phases of the transmission amplitudes is shown on the basis of our numerical calculations. In **Fig. 1** it is seen that, owing to the large value (~60) of the parameter $a$, introduced in Eq. (7), quite high nonlinearities appear even at the relatively moderate laser intensity (~$10^9$ W/cm$^2$) we are considering.

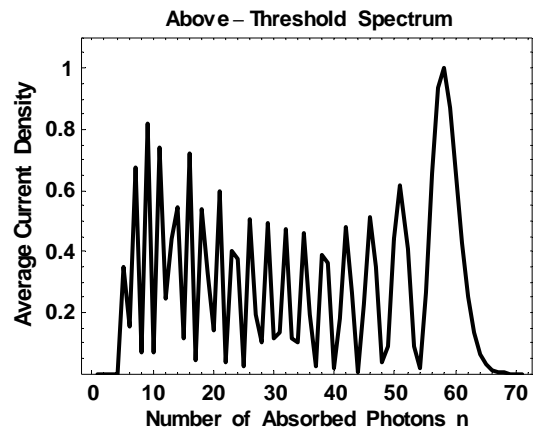

**Fig. 1** Shows the time-averaged spectrum of the above-threshold electron current density for the parameters introduced in the text above. The relative current density components $j_t(n) = (p_n/q_0)\cdot|T_n|^2$ of Eq. (8a) are normalized to their maximum value ~0.02. The discrete points are connected by thick lines in order to guide the eye.

The next figure shows the numerical results for the phase differences $\phi_n - \phi_{n+1}^*$ (mod $2\pi$) of the transmission amplitudes $T_n$ .

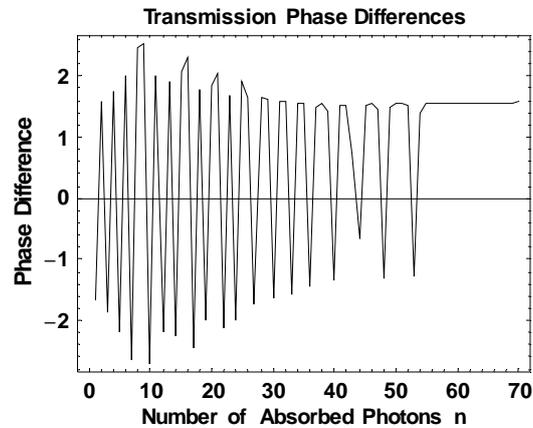

**Fig. 2** Shows the dependence of phase differences $\arg[T_{n+1}^* T_n]$ in radians on the number of absorbed photons $n$, corresponding to the spectrum shown in **Fig.1**. The discrete points are connected with thin lines in order to guide the eye. For relatively low-order processes the phase difference varies quite irregularly, but for the large energy wing, for $54 < n$, it is stationary, which means that these components are strictly locked, which is in complete accord with the first approximate formula in Eq. (9). On the basis of this figure we expect that the large-energy Floquet components interfere constructively, because of the regular behaviour of their phases.

The detailed structure of the spectrum and the associated phases around the maximum will be summarized on the next figure.

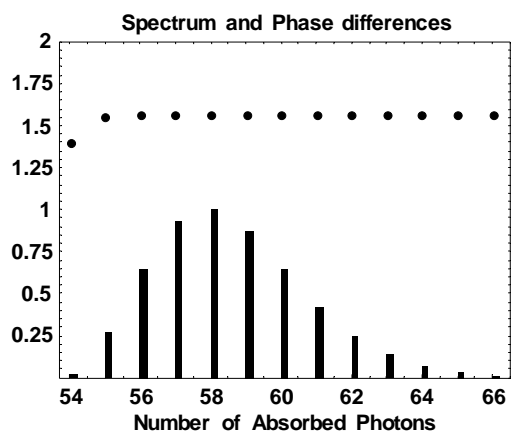

**Fig. 3** Shows the large energy wing of the above-threshold spectrum represented by vertical bars (normalized to their maximum value ~0.02. This is a magnification of the end of the spectrum shown already in **Fig. 1**, and shows also at the large-energy part of **Fig. 2** , the dependence of phase differences $\phi_n - \phi_{n+1}$ in radians on the number of absorbed photons $n$, at the end of the multiphoton spectrum. These phase differences are represented by points. According to the approximate formula given by Eq. (9) these differences $\arg[T_{n+1}^*] - \arg[T_n]$ are quiet uniformly equal about to $\pi/2 \approx 1.57$, as is also seen on the figure.

## 4. Detailed temporal behaviour of the above-threshold current density; attosecond electron pulse trains

The superposition of the fundamental solutions describing the outgoing above-threshold de Broglie waves, presented after Eq. (5b), can be brought to the form

$$\Psi_{II} = \exp\left[+i\frac{U_D}{\hbar\omega}\cos\omega t\right]\sum_n T_n \exp\left[-2\pi i\cdot\left(\frac{E_0}{\hbar\omega}+n\right)\frac{t}{T} + 2\pi i\cdot\sqrt{\frac{2mc^2}{\hbar\omega}\left(\frac{E_0}{\hbar\omega}+n\right)}\frac{z}{\lambda}\right], \qquad (10)$$

where we see that a natural unit of time is the optical period $T$ (namely 2.6 femtoseconds for a Ti:Sa laser), and the unit of length measuring the distance perpendicular from the metal surface is given as $\lambda\sqrt{\hbar\omega/2mc^2}$ , which, with a good accuracy, is about 1 nanometer in the present case. In **Fig. 4** we shall present the spatial development of the time-behaviour of the outgoing current density.

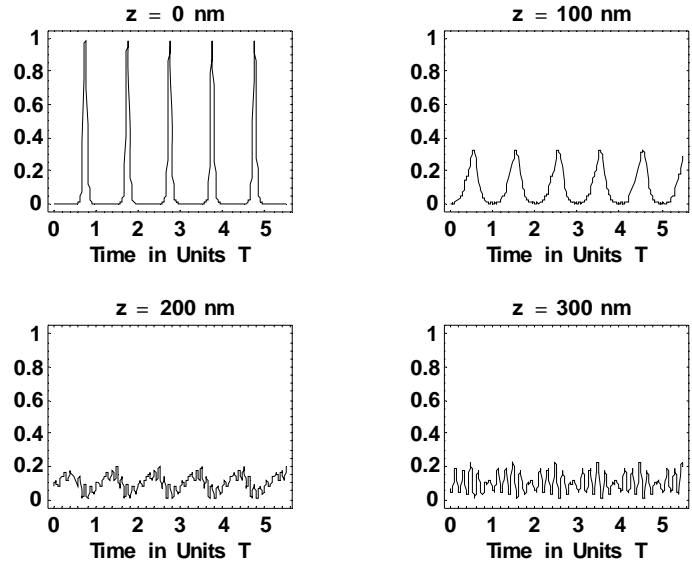

**Fig. 4** Shows the detailed time-behaviour of the coutgoing current density at different distances from the metal surface. Here we have plotted the dimensionless quantity $(\hbar / 4q_0)\,\mathrm{Re}[\Psi_{II}^{*}(-i\partial_z \Psi_{II})]$ by superimposing the large-energy components ($54 < n < 66$), whose phase are locked, according to **Fig. 3**. At $z = 0$ we received an ideal $T/10 \sim 250$ attosecond pulse train, which is gradually spoiled by the inherent dispersion of de Broglie waves propagating in vacuum (due to the non-trivial $z$-dependence of the phases), yielding a noise-like electron current density already at the distance of 300 nm from the metal.

The next figure summarizes the spatio-temporal behaviour of the outgoing current density.

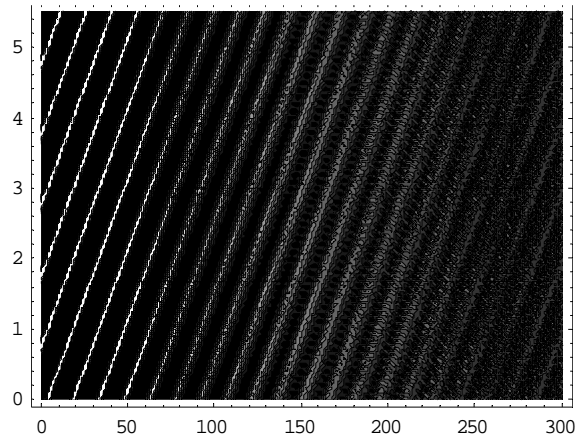

**Fig. 5** Shows the spatio-temporal behaviour of the outgoing current in the range {0 nm < $z$ < 300 nm , 0 $T$ < $t$ < 5.5 $T$, }. The white lines represent large values of the current density, on the other hand, the dark (grey or even black) regions indicate low values of the current density. By intersecting the figure vertically at some position $z$, and projecting the intersection to the time-axis (to the ordinate) we receive a representation of the time behaviour at that particular position. It is seen that the ideal attosecond pulse train at the immediate vicinity of the metal surface $z = 0$, represented by the light lines, is gradually washed out by the propagation from the surface. At $z = 300$ nm the temporal variation of the electron current density becomes already a noise-like background (represented by the dark band), as is also shown on the last figure in the graphics array of **Fig. 4**.

From the numerical study of contour plots of similar sort of **Fig. 5**, we have realized that revivals of the attosecond structure of the electron current density at very large (even at macroscopic) distances from the metal surface (e.g., we have checked in the spatial range 2 013 985 nm < z < 2 014 015 nm) can take place. The thickness of this "*revival layer*" is of order of 10 nm in this case. We have numerically checked for several ranges of the distances $z$ from the metal-vacuum interface, that the "revival layers" are separated usually by wider "*collaps bands*" where the electron signal has practically an irregular noise-like temporal behaviour. The next graphics array shows the temporal behavior of the current in the "collaps bands" and in the "revival layers" in a 555 nm range far from the metal-vacuum interface.

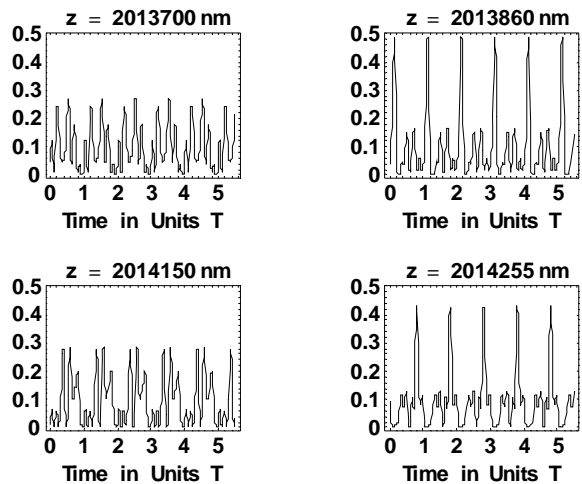

**Fig. 6** Illustrates the formation of the "revival layers" from the "collaps bands" in a 555 nm size range far from the metal surface at macroscopic distances (2 013 700 nm $< z <$ 2 014 255 nm). Here we have again plotted the dimensionless quantity $(\hbar / 4 q_0) \operatorname{Re}[\Psi_{II}^{*}(-i \partial_z \Psi_{II})]$ by superimposing the large-energy components ($54 < n < 66$), whose phases are locked, according to **Fig. 3**.

Since we have been discussing the relative current densities so far, at the end of the present Section we would like to make an estimate on the absolute size of the transmission current. According to the Fermi distribution of electrons in the metal, the total incoming current density impinging from inside to the metal-vacuum interface can be written as

$$J_{inco\min g} = (e \cdot n_e \cdot \upsilon_F) \cdot \left[ \frac{2}{(2\pi)^3} \cdot \frac{k_F^3}{n_e} \right] \cdot \int_0^{\infty} d\zeta \cdot \zeta \cdot \int_{-\infty}^{+\infty} d\xi \cdot \int_{-\infty}^{+\infty} d\eta \frac{1}{\exp[(E_F / kT) \cdot (\zeta^2 + \xi^2 + \eta^2 - 1)] + 1},$$

(11a)

where we have introduced the dimensionless variables

$$\xi \equiv q_x / m\upsilon_F \ , \ \ \eta \equiv q_y / m\upsilon_F \ \ \text{and} \ \ \ \zeta \equiv q_z / m\upsilon_F \ , \tag{11b}$$

and the Fermi wave number $k_F \equiv m\upsilon_F / \hbar = 1.21 \times 10^8 \, cm^{-1}$, with $\hbar$ denoting the Planck's constans $\hbar \equiv h / 2\pi$ divided by $2\pi$, as usual. The Fermi velocity $\upsilon_F = 1.40 \times 10^8 \, cm/s$ is related to the Fermi energy by the relation $E_F = m\upsilon_F^2 / 2 = 5.51 eV$. For a gold target the first prefactor yields $(e \cdot n_e \cdot \upsilon_F) = (1.6 \times 10^{-19} As) \cdot (5.9 \times 10^{22} cm^{-3}) \cdot (1.4 \times 10^8 cm/s) = 1.3216 \times 10^{12} A/cm^2$, because the electron density equals $n_e = 5.9 \times 10^{22} cm^{-3}$, according to Radzig and Smirnov (1985). The double integral with respect to $\xi$ and $\eta$ can be analytically performed, as is shown e.g. by Sokolov (1967),

$$\int\limits_{-\infty}^{+\infty} d\xi \cdot \int\limits_{-\infty}^{+\infty} d\eta \, \frac{1}{\exp[(E_F/kT) \cdot (\zeta^2 + \xi^2 + \eta^2 - 1)] + 1} = \frac{\pi}{(E_F/kT)} \cdot \log\Big\{ \, 1 + \exp[-(E_F/kT) \cdot (\zeta^2 - 1)] \, \Big\}$$

(11c)

Thus, the incoming current density coming from one cubic centimeter can be expressed as

$$J_{incoming} = (e \cdot n_e \cdot \upsilon_F) \cdot \left[ \frac{2}{(2\pi)^3} \cdot \frac{k_F^3}{n_e} \right] \cdot \frac{\pi}{(E_F/kT)} \int\limits_0^\infty d\zeta \cdot \zeta \cdot \log\Big\{ \, 1 + \exp[-(E_F/kT) \cdot (\zeta^2 - 1)] \, \Big\}$$

(11d)

Since $E_F / kT = 5.51 / 0.0236 \approx 233.5$ (where we have taken into account that $kT = 0.0236$ for $T = 273 \, K$), $[2/(2\pi)^3] \cdot (k_F^3 / n_e) = 0.2421$ and $[\pi/(E_F/kT)] = 0.01346$ we have for the incoming current density from one cubic centimeter,

$$J_{incoming} = [4.287 \times 10^9 \, A/cm^2] \cdot \int\limits_0^\infty d\zeta \cdot \zeta \cdot \log\Big\{ \, 1 + \exp[-233.5 \cdot (\zeta^2 - 1)] \, \Big\}, \tag{12}$$

where $\zeta \equiv q_{z0} / m\upsilon_F$ is the initial scaled momentum of the test electron. Of course, the interaction volume is not one cubic centimeter, but much smaller, of order of $\delta_{skin}\lambda^2 \approx (\omega/\omega_p)\lambda^3 = 8.82 \times 10^{-14} cm^3$, so the numerical prefactor should be reduced accordingly; $(8.82 \times 10^{-14}) \cdot [4.287 \times 10^9 \, A/cm^2] = 3.78 \times 10^{-4} \, A/cm^2$. Since the numerical value of the dimensionless integral in Eq. (12) is 58.3785, the total current density impinging perpendicularly on the metal-vacuum interface from the interior of the metal is about $2.265 \times 10^{-2}$ *A/cm²*. As we have seen above, the maximum value of the relative current density is of order of 2%, hence the value of the outgoing current density is expected at maximum a few tenth of *mA/cm²*, which is, on the other hand, still a relatively large value fot an electron source.

## 5. Summary

In the Introduction of the present paper we have given an overview on the theoretical and experimental research carried out recently on the ultrashort laser pulses with matter. We have also discussed the similarities and differences of the propagation properties of short light pulses and electron pulses. We would like to emphasize that in the scheme discussed in the present paper, the production of attosecond electron pulses is a result of interference of de Broglie waves of above threshold Fourier components of the total electronic wave with the frequency spacing $\omega$, the frequency of the incoming laser field. Thus the electron pulse train stems from the interference of the "frequency comb" of the above-threshold components. Of course, if an ultrashort (e.g. few-cycle) laser pulse interact with a target (e.g. an atom or a metal surface), then the electron respose (ionization or photoeffect) is expected to be very short, too. But this is not the case in our present discussion, since we assume relatively long (many-cycle) laser pulses of moderate intensity. Just this assumption allows us to use the Floquet analysis of the Schrödinger equation. In Section 2 we briefly described the construction of the laser-induced oscillating double-layer potential (energy) of a test electron scattered by the metal surface, and derived the basic wave function matching equation of the the scattered electron. Moreover, we have given an approximate analytic expression for the multiphoton scattering amplitudes valid for large final electron energies. This model serves as our basis for studying the high-order multiphoton photoelectric effect. In Section 3 we presented the results of the numerical solutions of the matching equations for the scattering amplitudes corresponding to the elementary *n*-photon absorption of the test electron. The time-averaged above-threshold current components and the phase differences of the multiphoton transmission amplitudes has been discussed (see **Fig. 1**). The maximum of the relative current density is about 2%. Though the phase differences are erratic for low orders (see **Fig.2**), it turned out that in the large-energy wing of the spectrum essentially perfect phase locking is present (see **Fig. 3**) In Section 4 we studied the detailed temporal behaviour of the total transmitted current which results from the superposition (interference) of the above-threshold de Broglie waves. It turned out that the ideal *attosecond electron pulse train* appearing at the immediate vicinity of the metal surface collapses to an almost noise-like signal by the inherent kinematic dispersion of the electron waves as they perpendicularly propagate from the surface (see **Fig.4** and **Fig. 5**). On the other hand, there are regions quite far from the surface (even at macroscopic distances, see **Fig. 6**) where the clean attosecond structure of the electron current revives within a spatial range of order 555 nm. The *attosecond light signals* stemming from a single atom propagate in vacuum *without changing shape* with an intensity distribution of the form $f(t - \xi/c)$, where $\xi$ is the propagation direction. On the other hand, the Fourier components of the electron pulses have a non-trivial spatially dependent phase $2\pi i (2mc^2/\hbar\omega)^{1/2}[(E_0/\hbar\omega) + n]^{1/2}(z/\lambda)$, which can drastically vary as a function of the distance $z$ from the metal-vacuum interface, resulting in the degradation of the originally clean attosecon pulse train. This is even so, if the phases of the transmission coefficients $T_n$ are stricky locked (as has been shown in **Fig.3** for the large-energy wing of the spectrum). At certain spatial region of $z$ the interference can be constructive, and in other ranges destructive. Accordingly, during the propagation of the electron signal from the metal surface there appear consecutively "*collapse bands*" (with destructive interference) and "*revival layers*" (with constructive interference) of a few or more nanometer thickness. By now, no approximate analytic formula has been found by us, which would locate the position and size of the revival layers and the collaps bands. To find such a formula is a subject of our ongoing research. Anyway, we have found by trial (by chance) the position of couple of these regions by using spatio-temporal contour plots (similar to **Fig. 5**) in several domains. At the end of Section 4, it has also been shown that , the value of the outgoing current density can be estimated at maximum a few tenth of $mA/cm^2$, which is still a relatively large value. This is why we think that the mechanism discussed in the present paper may serve as a basis for constructing good quality electron injectors for e.g. particle acceleration. The attosecond structure of the electron current can, perhaps be used for monitoring of  ultrafast relaxation processes in single atoms or in condensed matter.

**Acknowledgment**

This work has been supported by the Hungarian National Scientific Research Foundation, OTKA, Grant No. T048324. One of us (Varró, S.) thanks partial support from the COST P14 Action, Reference Code COST-STSM-P14-02507.

**Caption to Figure 1**

Shows the time-averaged spectrum of the above-threshold electron current density for the parameters introduced in the text above. The relative current density components $j_t(n) = (p_n / q_0) \cdot |T_n|^2$ of Eq. (8a) are normalized to their maximum value ~0.02. The discrete points are connected by thick lines in order to guide the eye.

**Caption to Figure 2**

Shows the dependence of phase differences $\arg[T_{n+1}^* T_n]$ in radians on the number of absorbed photons $n$, corresponding to the spectrum shown in **Fig.1**. The discrete points are connected with thin lines in order to guide the eye. For relatively low-order processes the phase difference varies quite irregularly, but for the large energy wing, for $54 < n$, it is stationary, which means that these components are strictly locked, which is in complete accord with the first approximate formula in Eq. (9). On the basis of this figure we expect that the large-energy Floquet components interfere constructively, because of the regular behaviour of their phases.

**Caption to Figure 3**

Shows the large energy wing of the above-threshold spectrum represented by vertical bars (normalized to their maximum value ~0.02. This is a magnification of the end of the spectrum shown already in **Fig. 1**, and shows also at the large-energy part of **Fig. 2**, the dependence of phase differences $\phi_n - \phi_{n+1}$ in radians on the number of absorbed photons $n$, at the end of the multiphoton spectrum. These phase differences are represented by points. According to the approximate formula given by Eq. (9) these differences $\arg[T_{n+1}^*] - \arg[T_n]$ are quiet uniformly equal about to $\pi / 2 \approx 1.57$, as is also seen on the figure.

**Caption to Figure 4**

Shows the detailed time-behaviour of the coutgoing current density at different distances from the metal surface. Here we have plotted the dimensionless quantity $(\hbar/4q_0)\,\mathrm{Re}[\Psi_{II}^*(-i\partial_z\Psi_{II})]$ by superimposing the large-energy components ($54 < n < 66$), whose phase are locked, according to **Fig. 3**. At $z = 0$ we received an ideal $T/10 \sim 250$ attosecond pulse train, which is gradually spoiled by the inherent dispersion of de Broglie waves propagating in vacuum (due to the non-trivial $z$-dependence of the phases), yielding a noise-like electron current density already at the distance of 300 nm from the metal.

**Caption to Figure 5**

Shows the spatio-temporal behaviour of the outgoing current in the range {0 nm $< z <$ 300 nm , 0 $T < t <$ 5.5 $T$, }. The white lines represent large values of the current density, on the other hand, the dark (grey or even black) regions indicate low values of the current density. By intersecting the figure vertically at some position $z$, and projecting the intersection to the time-axis (to the ordinate) we receive a representation of the time behaviour at that particular position. It is seen that the ideal attosecond pulse train at the immediate vicinity of the metal surface $z = 0$, represented by the light lines, is gradually washed out by the propagation from the surface. At $z = 300$ nm the temporal variation of the electron current density becomes already a noise-like background (represented by the dark band), as is also shown on the last figure in the graphics array of **Fig. 4**.

**Caption to Figure 6**

Illustrates the formation of the "revival layers" from the "collaps bands" in a 555 nm size range far from the metal surface at macroscopic distances (2 013 700 nm < $z$ < 2 014 255 nm). Here we have again plotted the dimensionless quantity $(\hbar/4q_0)\,\mathrm{Re}[\Psi_{II}^*(-i\partial_z\Psi_{II})]$ by superimposing the large-energy components (54 < $n$ < 66), whose phases are locked, according to **Fig. 3**.

**Figure 1** ("Varro_Elatto_Fig1.eps")

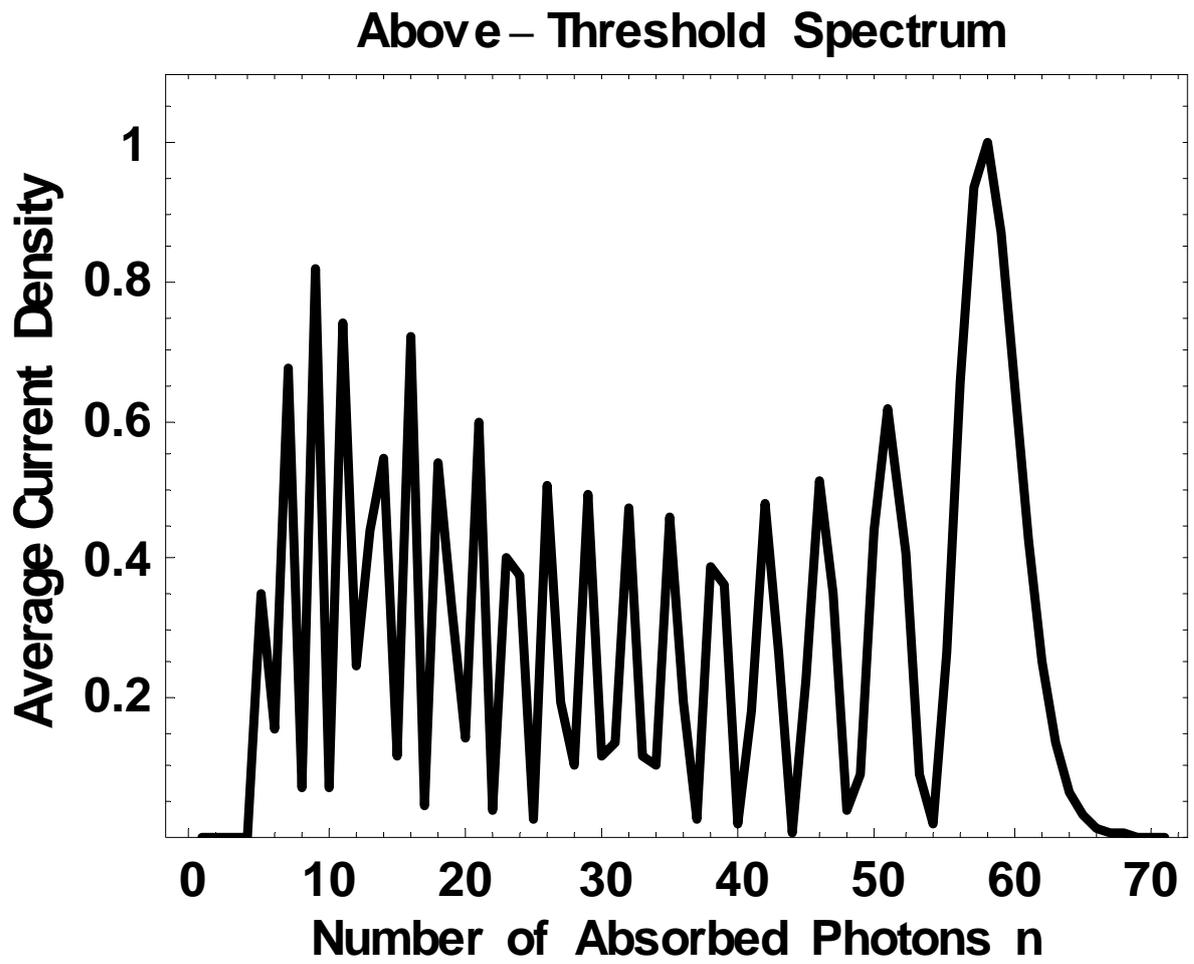

**Figure 2** ("Varro_Elatto_Fig2.eps")

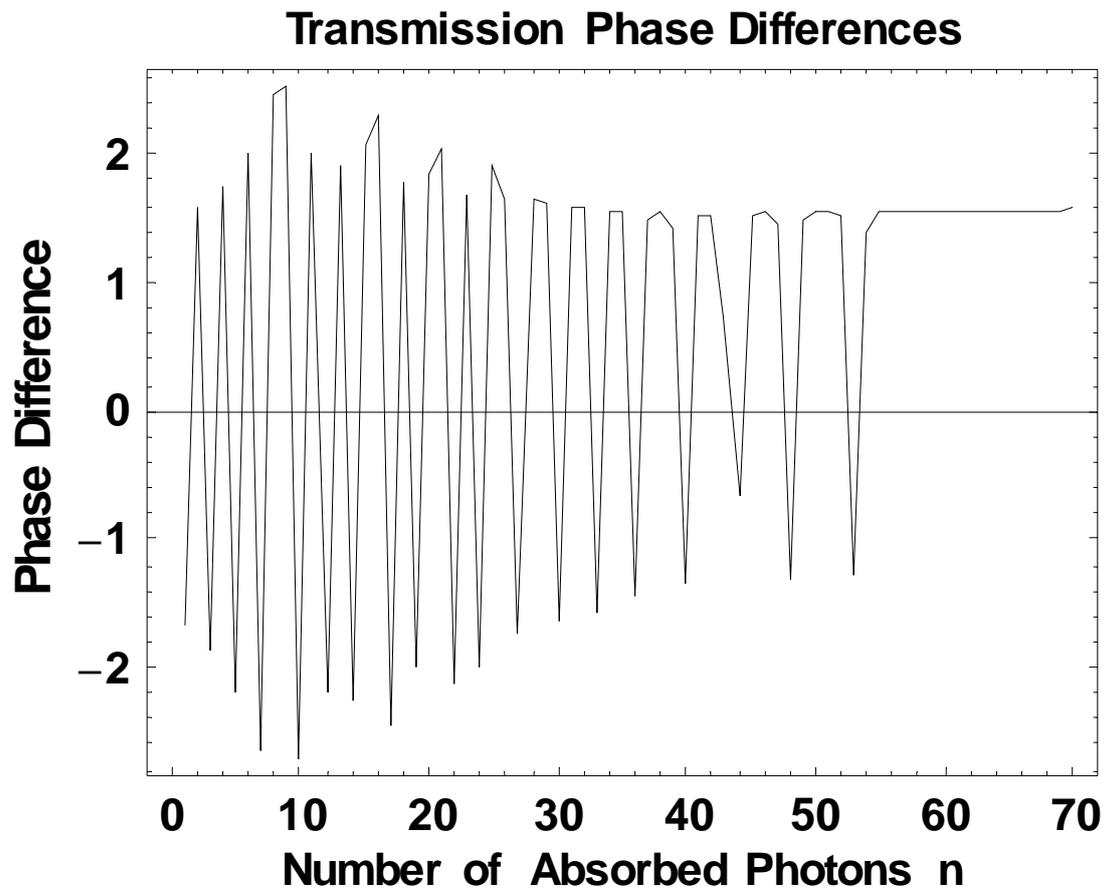

**Figure 3** ("Varro_Elatto_Fig3.eps")

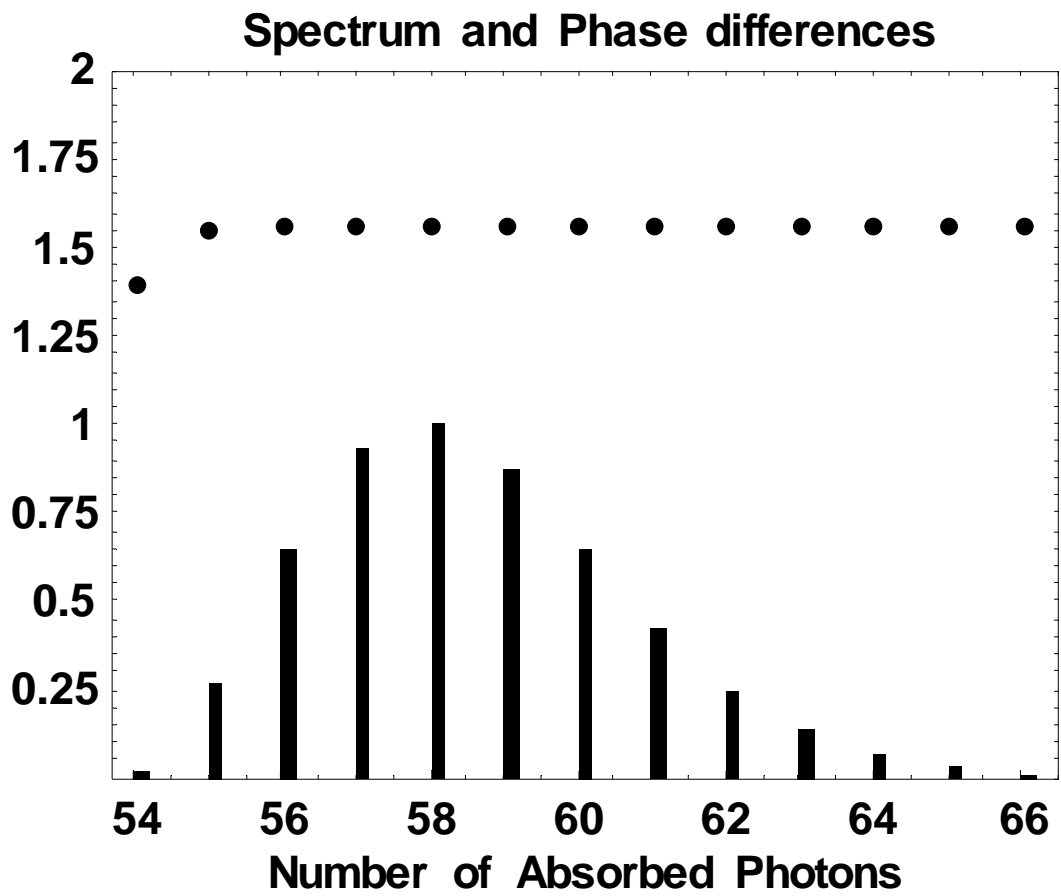

**Figure 4** ("Varro_Elatto_Fig4Array.eps")

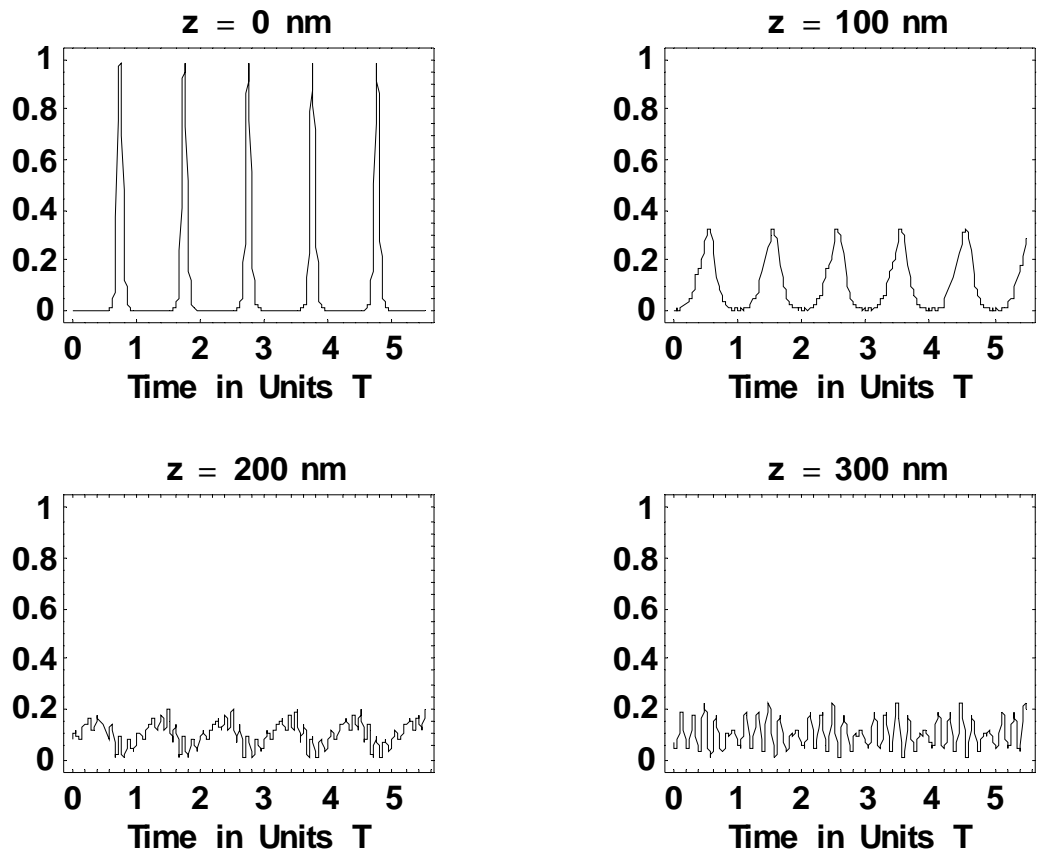

**Figure 5** (Varro_Elatto_Fig5Contour.eps")

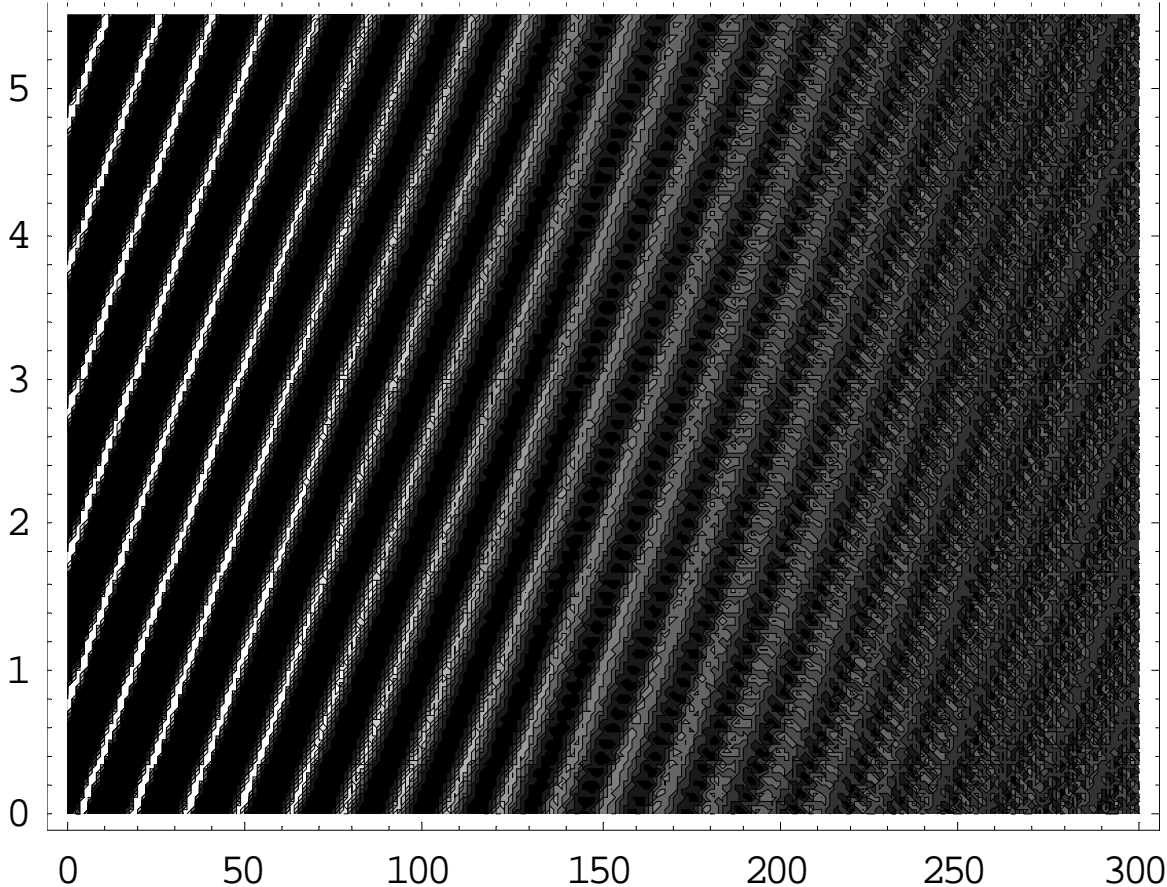

**Figure 6** (Varro_Elatto_Fig6Array.eps")

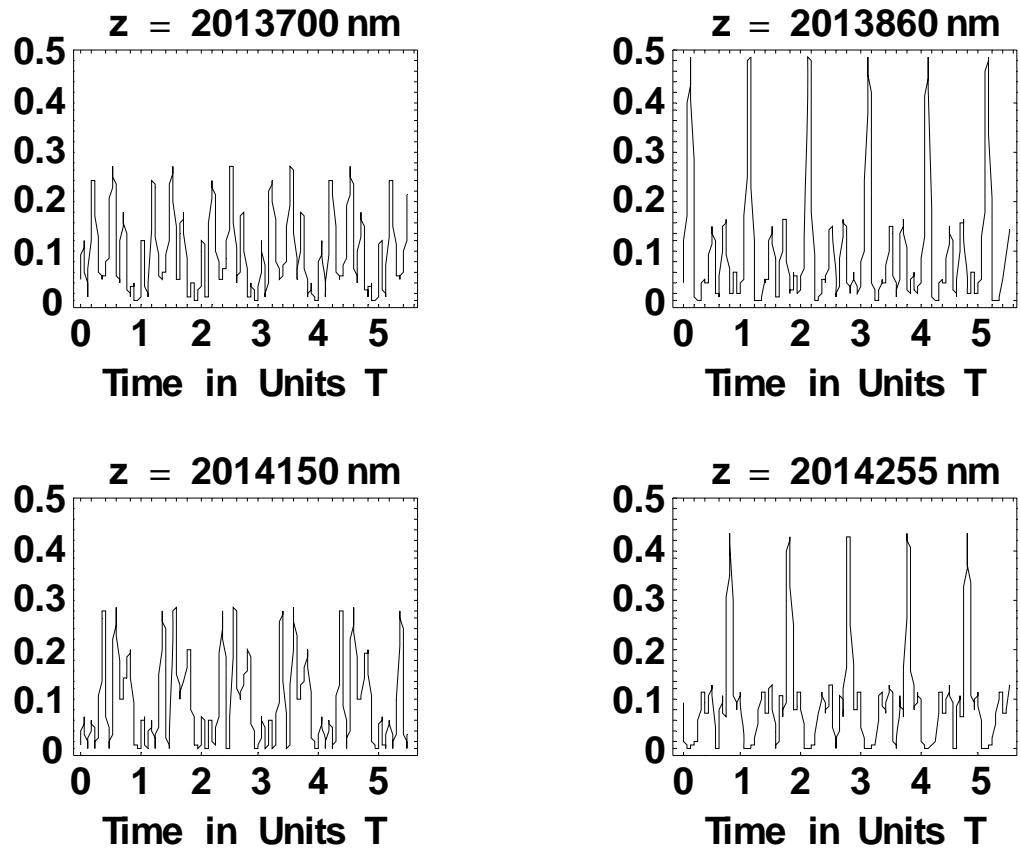